\documentclass[sigconf, nonacm]{acmart}

\AtBeginDocument{%
  \providecommand\BibTeX{{%
    \normalfont B\kern-0.5em{\scshape i\kern-0.25em b}\kern-0.8em\TeX}}}

\copyrightyear{2022}
\acmYear{2022}
\setcopyright{acmlicensed}
\acmConference[KDD '22]{Proceedings of the 28th ACM SIGKDD Conference on Knowledge Discovery and Data Mining}{August 14--18, 2022}{Washington, DC, USA}
\acmBooktitle{Proceedings of the 28th ACM SIGKDD Conference on Knowledge Discovery and Data Mining (KDD '22), August 14--18, 2022, Washington, DC, USA}
\acmPrice{15.00}
\acmDOI{10.1145/3534678.3539099}
\acmISBN{978-1-4503-9385-0/22/08}

\begin{document}

\title{Rapid Regression Detection in Software Deployments through Sequential Testing}


\author{Michael Lindon}
\affiliation{%
  \institution{Netflix, Inc}
  \city{Los Gatos, CA}
  \country{USA}}
\email{mlindon@netflix.com}

\author{Chris Sanden}
\affiliation{%
  \institution{Netflix, Inc}
  \city{Los Gatos, CA}
  \country{USA}}
\email{csanden@netflix.com}

\author{Vaché Shirikian}
\affiliation{%
  \institution{Netflix, Inc}
  \city{Los Gatos, CA}
  \country{USA}}
\email{vache@netflix.com}


\begin{abstract}
The practice of continuous deployment has enabled companies to reduce time-to-market by increasing the rate at which software can be deployed. However, deploying more frequently bears the risk that occasionally defective changes are released. For Internet companies, this has the potential to degrade the user experience and increase user abandonment. Therefore, quality control gates are an important component of the software delivery process. These are used to build confidence in the reliability of a release or change. Towards this end, a common approach is to perform a canary test to evaluate new software under production workloads. Detecting defects as early as possible is necessary to reduce exposure and to provide immediate feedback to the developer.

We present a statistical framework for rapidly detecting regressions in software deployments. Our approach is based on sequential tests of stochastic order and of equality in distribution. This enables canary tests to be continuously monitored, permitting regressions to be rapidly detected while strictly controlling the false detection probability throughout. The utility of this approach is demonstrated based on two case studies at Netflix.
\end{abstract}

\ccsdesc[500]{Mathematics of computing~Hypothesis testing and confidence interval computation}
\ccsdesc[500]{Software and its engineering~Software testing and debugging}
\ccsdesc[300]{Software and its engineering~Empirical software validation}
\ccsdesc[300]{Software and its engineering~Rapid application development}

\keywords{A/B testing; regression detection; canary release; sequential testing; experimentation; canary testing; software delivery; anytime-valid inference; confidence sequences}

\maketitle

\section{Introduction}
\label{sec:intro}
Many companies have invested in optimizing the software release process to make developers more agile and permit new features to reach users faster. Over the years, best practices have evolved to continuous and progressive delivery models, where software can be deployed at any time with a great deal of automation \cite{Humble10}. This has resulted in rapid development cycles where changes are rolled out constantly, dramatically reducing the time taken for new features to reach end users. However, these changes pose an inherent risk as it is inevitable that something will go wrong. For Internet companies such as Netflix, this has the potential to degrade the user experience and increase user abandonment \cite{Kohavi13}. 

While test suites can be successful in catching functional problems within internal test environments, performance regressions may, at times, only appear under production workloads \cite{Foo15}. Furthermore, it can be expensive to comprehensively test the quality and performance of software via internal testing \cite{Xia19}. Maintaining an integration and testing environment that fully replicates the production environment is challenging and perhaps impossible. There are likely to be many differences between the state of the test environment and the state of the production environment. Therefore, it's inevitable that software changes will make it past sophisticated test suites, enter into the production environment, and degrade the user experience. Quality gates \cite{Schermann16} within the software delivery process are therefore needed that are able to quickly and accurately detect defective software under production workloads.

\subsection{Regression-Driven Experiments}
To minimize the risk of deploying defective software, engineering practices have evolved towards regression-driven experimentation, namely canary testing \cite{Schermann18, Schermann17}. This practice involves exposing a change to a small group of users or traffic and studying its effects under production workloads \cite{Humble10}. This approach reduces the negative impact of defective software by first validating the change on a subset of the population. If no significant effects are observed, the change is then rolled out to the entire population.

One common approach to canary testing is to treat it as a controlled experiment \cite{kayenta, Davidovic18, Tarvo15}. In this approach, a small portion of users or traffic are randomly assigned to one of two variants: control or treatment. The treatment represents the release candidate and contains the change to be tested while the control is a copy of the previous version. This approach is related to an A/B test design found in online controlled experiments. The benefit of this approach is that it establishes a causal relationship between the software and a change in performance. 

A major shortcoming of existing approaches is the reliance on \textit{fixed-$n$} tests. These are statistical tests that provide statistical guarantees, such as type I (false detection) error control, when used \textit{strictly once}. Historically these were developed to possess optimal statistical properties in applications where there was only one possible time to perform the analysis, such as when a statistician is simply handed a dataset or when all observations arrive simultaneously. However, in a canary test, observations can arrive \textit{sequentially} which provide many opportunities for testing. Herein lies the difficulty of using fixed-$n$ tests in sequential applications; the choice of when to perform the test. Perform the test too early, and small regressions may go undetected (a high type II error probability). Perform the test too late and large regressions have been permitted to cause harm. Ideally one seeks to detect large regressions as soon as possible and end the test, preventing further experimental units from being exposed. In contrast to fixed-$n$ methodology, a natural form of scientific inquiry in sequential applications is to continue to collect data until a hypothesis has been proven or disproven.

Without sufficient statistical tooling, an erroneous practice of ``peeking'' has evolved where fixed-$n$ tests are repeatedly applied on an accumulating set of data \cite{peeking}. The problem with this practice is a lack of control over type I errors. One wants to apply the test frequently, to detect regressions as soon as possible, but the probability of making a type I error with a fixed-$n$ test increases with each usage \cite{armitage}. We posit that canary tests should always have a strict type I error guarantee. Otherwise, the experiment would be unreliable, producing many false alerts which can increase manual intervention and reduce delivery cadence. 

This issue can be resolved by the development of modern \textit{sequential} statistical tests, which are optimized for applications where data arrives sequentially. The type I error is controlled no matter how many times they are used, even after every datapoint if desired, allowing experiments to be continuously monitored. This provides the ability to perform optional stopping, stopping the experiment in a data-dependent way as soon as a hypothesis is rejected i.e. when a regression is detected. Towards this end, sequential tests for difference-in-means of Gaussian random variables have already been widely used for online A/B experiments \cite{Johari15,peeking,Zhao18}. However, we argue that performing inference about means is too limited for canary tests, for not all bugs or performance regressions can be captured by differences in the mean alone, as the following example demonstrates. Consider \textit{PlayDelay}, an important key indicator of the streaming quality at Netflix, which measures the time taken for a title to start once the user has hit the play button. It is possible for the mean of the control and treatment to be the same, but for the tails of the distribution to be heavier in the treatment, resulting in an increased occurrence of extreme values. Large values of \textit{PlayDelay}, even if infrequent, are unacceptable and considered a severe performance regression by our engineering teams. Differences in the mean are therefore too narrow in scope for defining performance regressions. 

\subsection{Contributions}
The contributions of this paper are to provide developers with a new statistical framework for the rapid testing of rich hypotheses in canary tests. We first extend the statistical definition of bugs and performance regressions beyond simple differences in the mean, and provide a decision-theoretic approach to reasoning about the control and treatment. We propose a sequential canary testing framework using sequential tests of equality and stochastic ordering between distributions to automate testing. These allow \textit{any difference} across the entire \textit{distribution}, not just the mean, to be tested in real-time and identified as soon as possible. This captures a much broader class of bugs and performance regressions, while providing strict type I error guarantees specified by the developer. 

Our statistical methodology is based on the confidence sequences and sequential tests of \citet{Howard21}. We build upon their tests by providing sequential $p$-values for testing equality and stochastic ordering among distributions, a complementary stopping rule for accepting approximately true null hypotheses, and an upper bound on the time taken for the test to stop. The contribution of sequential $p$-values is necessary if one wishes to measure the strength of evidence against the null hypothesis and also if one wishes to use multiple testing corrections to control the false discovery rate \cite{Johari15}. The complementary stopping rule is necessary if one wishes the test to stop in finite time.  Without such a stopping rule the test is "open-ended", meaning it only stops when the null is rejected. If the null is true, then with probability at least $1-\alpha$ the test would run indefinitely. In practise, developers need the canary test to finish in a finite time. Our contribution allows developer to stop the canary test and conclude with a high degree of confidence that any difference, if it exists, is not practically meaningful.

The complementary stopping rule also provides an upper bound on the maximum number of samples required by the test to reject the null or accept an approximate null at a user-specified tolerance and confidence level. This gives the developer a meaningful maximum number of samples required for the canary test and helps with planning. We also emphasize along the way that \textit{estimation} is just as valuable as \textit{testing} in this context. While the hypothesis testing component is useful for automating the logic that terminates the experiment, we stress the use of ``anytime-valid'' confidence sequences across all quantiles of each distribution. These can be visualized at any time, without any need to be concerned about peeking, and can be used by developers for anytime-valid insights into what would otherwise be a black box. This gives a precise description of the differences between control and treatment and aids in developer \textit{learnings}.

This paper is organized as follows. In Section \ref{related work} we present related work. Section \ref{methodology} introduces the statistical methodology, first by developing the mathematics for the fixed-$n$ case, and then generalizing it to work sequentially. Section \ref{sec:casestudies} demonstrates the utility of this approach based on two case studies from Netflix. In Section \ref{sec:discussion} we reflect on our learnings and conclude the paper in Section \ref{sec:conclusions}.

\section{Related Work} 
\label{related work}
Canary testing systems \cite{Tarvo15, Davidovic18} have been proposed for developers who wish to conduct regression-driven experiments. These systems are based on the design of a controlled experiment and rely upon statistical tests that provide a type I error guarantee when used exactly once. Due to this, these systems can suffer from the erroneous practice of “peeking” if developers repeatedly apply the statistical tests in an attempt to detect regressions early. 

At Netflix, \textit{Kayenta} \cite{kayenta} has been used extensively for canary testing. Similar to \cite{Tarvo15, Davidovic18}, this approach is based on the design of a controlled experiment. To evaluate the outcome of the canary test, Kayenta uses fixed-$n$ statistical tests such as the Mann-Whitney U test. While these tests should be used strictly once we have found that developers will perform the tests multiple times during an experiment in an attempt to detect regressions early. This motivated our investment into sequential testing.

The statistics literature on sequential testing dates back at least to \cite{Wald45} with the introduction of the mixture sequential probability ratio test (mSPRT). Introductory texts to the subject can be found in \cite{Waldbook, Siegmund85}. Johari et al. \cite{Johari15, peeking} proposed an ``anytime-valid'' sequential inference framework for differences in the means of Gaussian random variables using the mSPRT to provide confidence sequences and sequential $p$-values. In addition to being used in commercial A/B testing software, \cite{Zhao18} use this framework for managing the automated rollout of new software features. However, they formulate performance regressions as differences in the mean which are too narrow in scope to capture the breadth of potential problems.

Our approach is based on the results of Howard et al. \cite{Howard21}. The authors provide confidence sequences that hold uniformly over time across all quantiles of a distribution, which they use in the appendix to construct sequential tests of equality in distribution and stochastic dominance. 

\section{Statistical Methodology} 
\label{methodology}
In this section, we first define a performance regression in statistical terms. Then, we will describe the sequential test. For brevity, we'll refer to the control and treatment simply as arms A and B of the experiment.

\subsection{Beyond Inference on the Mean}
\label{sec:beyondmean}
A canary testing system aims to identify bugs and performance regressions by observing changes in the distribution of key metrics between arms. There is a strong concern that small performance regressions accumulate and substantially degrade the user experience over time. Usually, there is a directional notion of ``desirable'' and ``undesirable'' changes in the distribution of a metric. Consider, for example, \textit{PlayDelay}. \textit{PlayDelay} is the time taken, in milliseconds, for a title to begin streaming after being requested by the user. A shift in the distribution of \textit{PlayDelay} toward larger values results in a poorer streaming experience, whereas a shift toward smaller values could be considered an improvement.

Many of the earlier works formulate performance regressions as differences in the mean between arms A and B \cite{Zhao18}, with a performance regression occurring if the mean shifts in the undesirable direction of that metric. We argue from a decision-theoretic perspective that comparing means alone is insufficient to define a performance regression. Consider two users, one with a fast internet connection, the other with a slow internet connection, and consider the effect of \textit{PlayDelay} on their satisfaction with the service. The former user is less affected by increases in \textit{PlayDelay}, as the resulting values are likely still small and manageable, but the latter user is more affected, as their values were already high to begin with. As streaming performance is correlated with user dissatisfaction, increases in \textit{PlayDelay} increase the risk of the latter user churning more than they do for the former. In other words, increases in \textit{PlayDelay} are not nearly as important for the average user as they are for users already at risk, and increases in the mean are not as important as increases in the tail of the distribution.

We have found the concept of \textit{stochastic ordering} helpful in elevating directional comparisons beyond comparisons of the mean. Let $A$ and $B$ denote single observations from arms A and B, respectively. Let $L$ denote a loss function that defines the loss associated with the value of an observation. A developer will then prefer arm B over arm A if 
\begin{equation}
    r_b - r_a = \mathbb{E}[L(B)]-\mathbb{E}[L(A)] \leq 0,
\end{equation}
i.e.\ if the expected loss (risk) is lower for arm B than it is for A. If the loss is linear, $L(x) = kx$, the arm with the smaller risk is simply the arm with the smaller mean. As illustrated with \textit{PlayDelay}, however, many practical loss functions are not linear. Increases of $\Delta$ in \textit{PlayDelay} are worse for users with already large values i.e. $L(b+\Delta) - L(b) > L(a+\Delta)-L(a)$ for $b>a$. This implies that the loss function is not linear but convex, and clearly illustrates why defining regressions in terms of the mean alone is insufficient. Fortunately, $r_b \leq r_a $ can be guaranteed for all nondecreasing loss functions $L$ if $B$ is \textit{stochastically less than or equal to} $A$.

Let $A$ and $B$ be random variables with distributions $F_a$ and $F_b$, respectively. $B$ is said to be stochastically less than or equal to $A$, written $B\preccurlyeq A$, if $F_b(x) \geq F_a(x)$ $\forall x \in \mathcal{X}$. If, in addition, $F_b(x) > F_a(x)$ for some $x \in \mathcal{X}$, then $B$ is said to be \textit{strictly} stochastically less than $A$. This condition is also known as first-order stochastic dominance. The practical significance behind stochastic ordering is that $B \preccurlyeq A$ if and only if $\mathbb{E}[L(B)] \leq \mathbb{E}[L(A)]$ for all non-decreasing loss functions $L$ \cite{stochasticdominance}.
 
The previous result is helpful because it removes the burden of fully specifying a developer's subjective loss function $L$.  While different developers may disagree about the exact functional form of $L$, all agree that this function is non-decreasing because increases in the distribution toward larger values are undesirable. If all developers agree that decreases in the distribution toward smaller values are undesirable, then all agree that the loss function is non-increasing. In the end, the exact values of $r_b$ and $r_a$ are not specifically of interest. It is only relevant to know if $r_b \leq r_a$ and vice versa. 

To give a concrete example, let us return to studying the distribution of \textit{PlayDelay} in a canary test. Developers would like to catch ``increases'' in \textit{PlayDelay} in the release candidate, while no action is necessary for ``decreases''. Therefore, our null hypothesis is that observations from arm B are stochastically less than or equal to observations from arm A, $B\preccurlyeq A$, while the alternative is the complement of this, which we can summarize as
\begin{equation}
    \begin{split}
        H_0: F_b(x)&\geq F_a(x)  \hspace{1cm}\forall x \in \mathcal{X},\\
        H_1: F_b(x) &< F_a(x) \hspace{1cm} \text{for some } x \in \mathcal{X}.\\
    \end{split}
    \label{eq:onesided}
\end{equation}
Note that the complement of the null hypothesis is not $A\preccurlyeq B$ as stochastic ordering is only a partial ordering. If the null is indeed rejected, it is the responsibility of the developer to use their judgement.
Metrics for which ``decreases'' are bad can be tested similarly by replacing $\geq$ and $<$ with $\leq$ and $>$ respectively.

In many cases, the release candidate is not expected to have an effect on metrics at all. If the developer does not specify a direction, or one is not available from the application domain, then we test for \textit{any difference in the distribution}
        
        \begin{equation}
    \begin{split}
        H_0: F_a(x) &= F_b(x) \hspace{1cm}\forall x \in \mathcal{X},\\
        H_1: F_a(x) &\neq F_b(x) \hspace{1cm} \text{for some } x \in \mathcal{X},\\
    \end{split}
    \label{eq:twosided}
\end{equation}

Although the decision for a canary test to succeed or fail is formulated in terms of a hypothesis test, we stress that an equally important objective of a canary test is estimation, that is, about learning the quantile functions of arms A and B.

\subsection{Fixed-$n$ Inference}
\label{sec:fixed-n}
To build up to our full inferential procedure, we first consider the simpler fixed-$n$ case and then generalize it to the sequential case in Section \ref{sec:sequential}. The procedure is quite intuitive: estimate the distribution and/or quantile functions and then use these estimates to test hypotheses. For example, one could test the null hypothesis $F_a = F_b$ by asking if the confidence bands on $F_a$ and $F_b$ fail to intersect.
\subsubsection{One-Sample Distribution Function Confidence Band}
Consider first the Dvoretzky–Kiefer–Wolfowitz (DKWM) inequality \cite{dvoretzky, massart}
\begin{equation}
    \mathbb{P}[\|F_n-F\|_{\infty} > \varepsilon] \leq 2e^{-2n\varepsilon^2},
\label{eq:dkwinequality}
\end{equation}
where $F_n(x)=\frac{1}{n} \sum_{i=1}^{n}1[x_i\leq x]$ is the empirical distribution function and $\|\cdot\|_{\infty}$ is the sup-norm. It follows directly that a confidence band on the distribution function with coverage probability at least $1-\alpha$ can be constructed with
\begin{equation}
    \mathbb{P}[F_n^l(\alpha , x) \leq F(x) \leq F_n^u(\alpha, x) \,\forall x \in \mathcal{X}] \geq 1-\alpha,
    \label{eq:cdfband}
\end{equation}
where
\begin{equation}
\begin{split}
F_n^u(\alpha,x) &= \min(1,F_n(x)+\epsilon_n(\alpha)),\\
F_n^l(\alpha,x) &= \max(0,F_n(x)-\epsilon_n(\alpha)),\\
\epsilon_n(\alpha) &= \sqrt{\frac{\log \frac{2}{\alpha}}{2n}}
\end{split}
\label{eq:dkwepsilon}
\end{equation}
(for details see example 2.2.4.\ from \cite{bickel}).
\subsubsection{One-Sample Quantile Function Confidence Band}
 Alternatively, one can construct a confidence band on the quantile function instead, via
\begin{equation}
    \mathbb{P}[Q_n^{l}(\alpha, p) \leq Q(p) \leq Q_n^u(\alpha, p) \,\forall p \in [0,1]] \geq 1-\alpha,
    \label{eq:quantileband}
\end{equation}
where 
\begin{equation*}
\begin{split}
Q_n^u(\alpha, p) &= Q_n(p+\epsilon_n(\alpha)),\\
Q_n^l(\alpha, p) &= Q_n^\rightarrow (p-\epsilon_n(\alpha)),\\
Q_n(p) &= \sup\lbrace x \in \mathcal{X} : F_n(x) \leq p \rbrace,\\
Q_n^\rightarrow(p) &= 
\begin{cases}
    \sup\lbrace x \in \mathcal{X} : F_n(x) < p \rbrace& \text{if } p > 0\\
    \inf \lbrace x \in \mathcal{X} \rbrace              & \text{otherwise}.
\end{cases}
\end{split}
\end{equation*}
$Q_n(p)$ is the upper empirical quantile function (the right continuous inverse of the empirical distribution function and equal to the $\lfloor np \rfloor + 1$ order statistic of the data). $Q_n^\rightarrow (p)$ is the lower quantile function (the left continuous inverse of the empirical distribution function and equal to the $\lceil np \rceil$ order statistic of the data), with the definition extended to be  $\inf \lbrace x \in \mathcal{X} \rbrace   $ when $p\leq 0$.

\subsubsection{Two-Sample Distribution/Quantile Function Confidence Bands}
\begin{figure}[h]
\includegraphics[width=8cm]{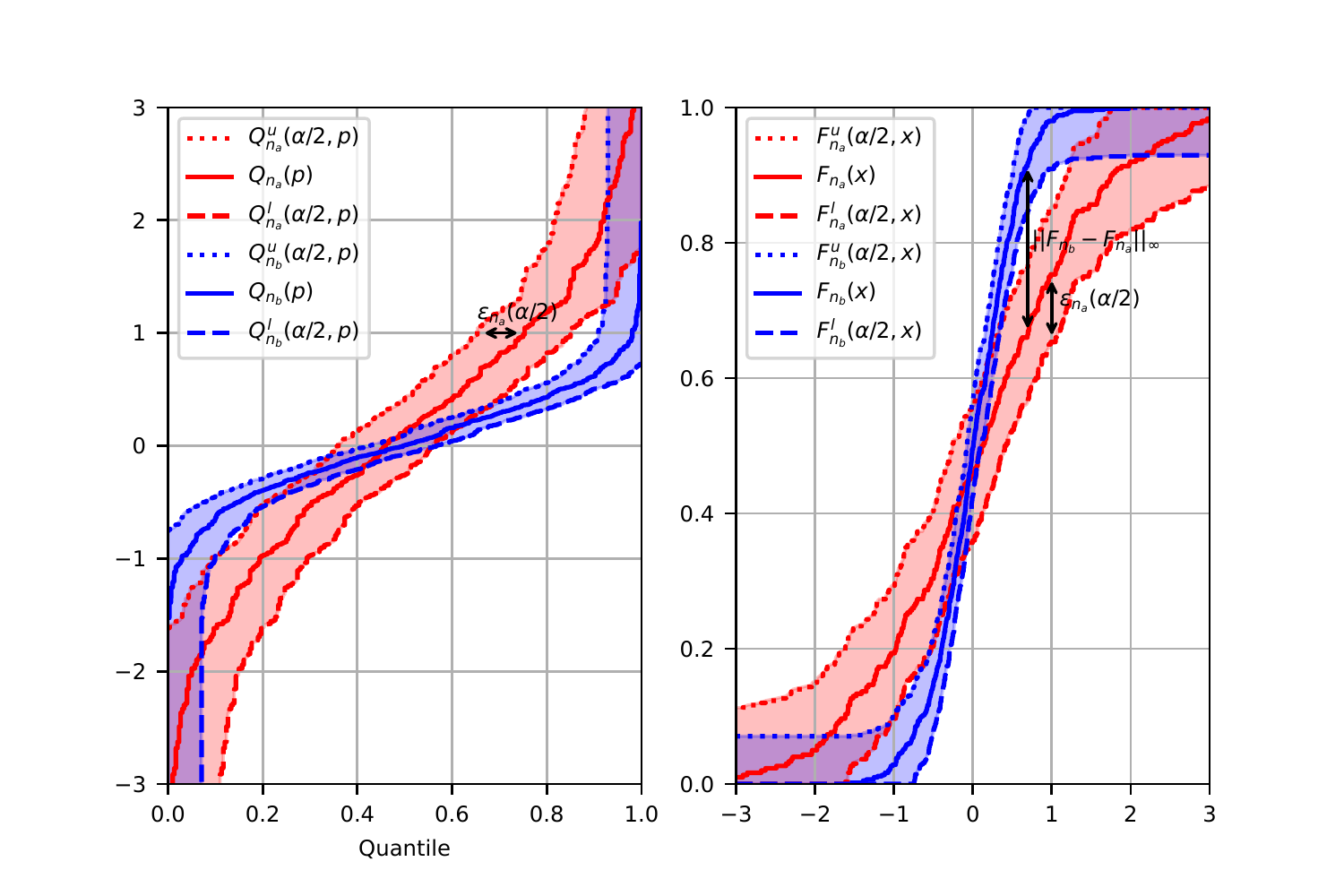}
\caption{Empirical quantile (left) and distribution (right) functions with $1-\alpha/2$ confidence bands from equations \eqref{eq:cdfband} and \eqref{eq:quantileband}, based on $n_a=300$ and $n_b=600$ samples from $\text{Normal}(0, 2)$ and $\text{Normal}(0,0.25)$ distributions for arms A and B respectively.}
\label{fig:dkwdistquants}
\end{figure}

Suppose one has two sets of i.i.d.\ samples resulting from an A/B test. Let the distribution function, empirical distribution function, sample size and sample from arm A be denoted $F_a$, $F_{n_a}$, $n_a$ and $x_1, x_2, \dots, x_{n_a}$ respectively. Similarly let the distribution function, empirical distribution function, sample size and sample from arm B be denoted $F_b$, $F_{n_b}$ $n_b$ and $y_1, y_2, \dots, y_{n_b}$ respectively. To obtain confidence bands on the distribution functions of arm A \textit{and} arm B that hold simultaneously with probability at least $1-\alpha$, then one can apply a union bound, yielding

\begin{equation}
\begin{split}
      \mathbb{P}[
      &F_{n_a}^l(\alpha/2 , x) \leq F_a(x) \leq F_{n_a}^u(\alpha/2, x) \\      
      &F_{n_b}^l(\alpha/2 , x) \leq F_b(x) \leq F_{n_b}^u(\alpha/2, x) \\
      &\forall x \in \mathcal{X}] \geq 1-\alpha,  
\end{split}
    \label{eq:twocdfband}
\end{equation}
The confidence bands for the quantile functions are handled similarly, and both are illustrated in \ref{fig:dkwdistquants}.

\subsubsection{Confidence Sets on the Difference}
Ultimately we would like to test hypotheses about the difference between $F_b$ and $F_a$, and so in this section we derive useful confidence statements about the difference.
 This section provides a confidence band on the function $d_{a,b} := F_b - F_a$ and a confidence interval for $\|d_{a,b}\|_\infty$. Figures that complement figure \ref{fig:dkwdistquants} in visualizing the difference between distribution functions are provided in Appendix \ref{app:absdiff}.

\begin{proposition}
Let the difference between distribution and empirical distribution functions be denoted  $d_{a,b} := F_b - F_a$ and $d_{n_a,n_b} := F_{n_b} - F_{n_a}$, respectively, with $F_{n_a}^{l}$, $F_{n_a}^{u}$, $F_{n_b}^{l}$, $F_{n_b}^{u}$ defined as in equation \eqref{eq:twocdfband}, then
\begin{equation}
     \mathbb{P} [d_{n_a,n_b}^l(\alpha,x)  \leq d_{a,b}(x) \leq d_{n_a,n_b}^u(\alpha,x) \forall x \in \mathcal{X}] \geq 1-\alpha,
\label{eq:diffconfband}
\end{equation}
where
\begin{equation*}
\begin{split}
d_{n_a,n_b}^u(\alpha,x) &= F_{n_b}^{u}(\alpha/2,x) -F_{n_a}^{l}(\alpha/2,x)\\
 d_{n_a,n_b}^l(\alpha,x)&= F_{n_b}^{l}(\alpha/2,x) -F_{n_a}^{u}(\alpha/2,x).\\
\end{split}
\end{equation*}
\end{proposition}
The confidence band on $d_{a,b}$ is an immediate consequence of the confidence bands on $F_a$ and $F_b$.
\begin{corollary}
\label{cor:infsupci}
\begin{equation*}
    \mathbb{P}\left[\underset{x\in\mathcal{X}}{\sup} d_{a,b}(x) \in \left[
    \underset{x\in\mathcal{X}}{\sup}d_{n_a,n_b}^{l}(\alpha, x), \underset{x\in\mathcal{X}}{\sup}d_{n_a,n_b}^{u}(\alpha,x)\right]\right] \geq 1-\alpha
\end{equation*}
\begin{equation*}
    \mathbb{P}\left[\underset{x\in\mathcal{X}}{\inf} d_{a,b}(x) \in \left[
    \underset{x\in\mathcal{X}}{\inf}d_{n_a,n_b}^{l}(\alpha, x), \underset{x\in\mathcal{X}}{\inf}d_{n_a,n_b}^{u}(\alpha,x)\right]\right] \geq 1-\alpha
\end{equation*}
\end{corollary}

\begin{corollary}
    \begin{equation}
     \mathbb{P}\left[
     \|d_{a,b}\|_\infty \in \left[ l_{n_a, n_b}(\alpha),
     u_{n_a, n_b}(\alpha) \right]\right] \geq 1-\alpha,
\label{eq:supnormci}
\end{equation}
where
\begin{equation*}
\begin{split}
    l_{n_a, n_b}(\alpha) &= \max \left(
    \left| \underset{x\in\mathcal{X}}{\inf}d_{n_a,n_b}^{u}(\alpha, x)\right|,\left|\underset{x\in\mathcal{X}}{\sup}d_{n_a,n_b}^{l}(\alpha,x)\right|  \right)\\
    u_{n_a, n_b}(\alpha) &= \max \left(
    \left| \underset{x\in\mathcal{X}}{\inf}d_{n_a,n_b}^{l}(\alpha, x)\right|,\left|\underset{x\in\mathcal{X}}{\sup}d_{n_a,n_b}^{u}(\alpha,x)\right|  \right)
\end{split}
\end{equation*}
\end{corollary}
\subsubsection{A Test of $A\preccurlyeq B$}
\begin{theorem}
\label{thm:a<=b}
The null hypothesis $F_a(x) \geq F_b(x) \, \forall x \in \mathcal{X}$ can be rejected at the $\alpha$-level if
\begin{equation*}
    \underset{x \in \mathcal{X}}{\sup d_{n_a,n_b}^{l}}(\alpha, x) > 0
\end{equation*}
The $p$-value for this test, $p_{n_a, n_b}^{\preccurlyeq}$, is the root of 
\begin{equation}
    f(\alpha) =  \|d_{n_a,n_b}^{+}\|_\infty - \epsilon_{n_a,n_b}(\alpha),
    \label{eq:dkwalbroot}
\end{equation}
where $d_{n_a,n_b}^{+}(x) = \max (d_{n_a,n_b}(x), 0)$ and  $\epsilon_{n_a,n_b}(\alpha) = \epsilon_{n_a}(\alpha/2)+ \epsilon_{n_b}(\alpha/2)$.
\end{theorem}
The proof is provided in Appendix \ref{app:a<=b}.
 When $n_a = n_b = n$, the root of equation \ref{eq:root} can be computed analytically as
\begin{equation}
 p_{n_a, n_b}^{\preccurlyeq} = 2e^{-\frac{n \|d_{n_a,n_b}^{+}\|_\infty^2}{2}}.
    \label{eq:dkwalbpvalue}
\end{equation}
When $\|d_{n_a,n_b}^{+}\|_\infty$ is large the $p$-value is small.
 When $n_a \neq n_b$ the root of equation \eqref{eq:dkwalbroot} must be found numerically. Some ``bracketing'' root-finding algorithms require the specification of an interval in which to seek the root, such as the bisection method \cite{bisection}. To construct lower and upper bounds on the $p$-value, define
\begin{equation}
    \label{eq:dkwanalyticroot}
    p_n(d) := 4e^{-\frac{n d^2}{2}}.
\end{equation}
 The $p$-value is lower-bounded by $p_{\max(n_a, n_b)}(\|d_{n_a,n_b}^{+}\|_\infty)$, and upper-bounded by $p_{\min(n_a, n_b)}(\|d_{n_a,n_b}^{+}\|_\infty)$.
 \begin{corollary}
     The null hypothesis $F_a(x) \leq F_b(x) \, \forall x \in \mathcal{X}$ $(A \succcurlyeq B)$ can be rejected at the $\alpha$-level if
\begin{equation*}
    \underset{x \in \mathcal{X}}{\inf} d_{n_a,n_b}^{u}(\alpha, x) < 0
\end{equation*}
The $p$-value for this test, $p_{n_a, n_b}^{\succcurlyeq}$, is the root of 
\begin{equation}
    f(\alpha) =  \|d_{n_a,n_b}^{-}\|_\infty - \epsilon_{n_a,n_b}(\alpha),
\end{equation}
where $d_{n_a,n_b}^{-}(x) = \max (-d_{n_a,n_b}(x), 0)$.
 \end{corollary}
\subsubsection{A Test of $A \overset{d}{=} B$}
\label{sec:dkwpvalue}
The null hypothesis $A \overset{d}{=} B$ is the intersection of the hypotheses $A \preccurlyeq B$ and $A \succcurlyeq B$ i.e.\ $F_b(x) \leq F_a(x)$ and $F_a(x) \leq F_b(x)$ for all $x\in \mathcal{X}$. This means that the results of the previous section can be re-purposed. 
The null hypothesis can be rejected at the $\alpha$-level if $\sup d_{n_a,n_b}^l(\alpha,x) > 0$ or $\inf d_{n_a,n_b}^u(\alpha,x) < 0$. 
Similarly, the $p$-value can be defined as $p_{n_a,n_b} = \min (p_{n_a,n_b}^\preccurlyeq, p_{n_a,n_b}^\succcurlyeq)$. 
When $n_a = n_b = n$, $p_{n_a,n_b} = 4 \exp(-n/2 \|d_{n_a,n_b}\|_\infty^2)$, otherwise it is sandwiched between $p_{\max(n_a, n_b)}(\|d_{n_a,n_b}\|_\infty)$ and $p_{\min(n_a, n_b)}(\|d_{n_a,n_b}\|_\infty)$.

To gain some intuition for the mathematics, it is useful to recognize that the following are equivalent

\begin{itemize}
    \item The $1-\alpha$ lower confidence bound for $\|d_{a,b}\|_\infty$, $l_{n_a,n_b}(\alpha)$, is greater than zero
    \item Either $\sup d_{n_a,n_b}^l(\alpha,x) > 0$ or $\inf d_{n_a,n_b}^u(\alpha,x) < 0$
    \item The $1-\alpha$ confidence band for the difference function $d_{a,b}$ in equation \eqref{eq:diffconfband} excludes 0 for some $x \in \mathcal{X}$
    \item The $1-\alpha/2$ confidence bands for the distribution functions in equation \eqref{eq:twocdfband} fail to intersect for some $x\in\mathcal{X}$
    \item The $1-\alpha/2$ confidence bands on the quantile functions fail to intersect for some $p\in [0,1]$
\end{itemize}
These statements are visualized with the help of figures \ref{fig:dkwdistquants} and \ref{fig:dkwdiff}. 

\subsubsection{Sample Size Calculations}
\label{sec:samplesize}
A sample size calculation can be performed to obtain a $1-\alpha$ confidence band for $d_{a,b}$ of a desired radius.
 When $n_a = n_b = n$ a radius of at most $r$ can be achieved with
\begin{equation}
    n = 2 \frac{\log\left(\frac{4}{\alpha}\right)}{r^2}.
\end{equation}
A choice of $r$ can help reason if the difference is practically meaningful. Suppose one considers differences between distribution functions less than $\tau$ to be practically irrelevant. By choosing $r \leq \tau/2$ the diameter of the confidence band is less than $\tau$. If the confidence band contains 0 for all $x \in \mathcal{X}$, then $\sup d_{n_a, n_b}^{u}(\alpha,x) \leq \tau$ and $\inf d_{n_a,n_b}^{l}\geq -\tau$, which implies $u_{n_a, n_b}(\alpha)\leq \tau$. One can conclude with confidence $1-\alpha$ that $\|d_{a,b}\|_\infty \leq \tau$.

\subsection{Sequential Inference}
\label{sec:sequential}
The aforementioned coverage and type I error probabilities of the statistical procedure constructed from the DKWM inequality hold when the analysis is performed only once. This is an example of a fixed-$n$ test which is appropriate when the analysis is to be performed only once. This test is, however, not appropriate when data arrives sequentially and the analysis is intended to be performed continuously, as is the case with canary testing. To allow continuous monitoring we seek a \textit{confidence sequence} 
\begin{equation}
    \mathbb{P}[F_n^l(\alpha , x) \leq F(x) \leq F_n^u(\alpha, x) \,\forall x \in \mathcal{X}\, \forall n \in \mathbb{N}] \geq 1-\alpha.
    \label{eq:timeuniformcdfband}
\end{equation}
This extends the confidence band in \eqref{eq:cdfband} to hold for all $n \in \mathbb{N}$. This, in turn, allows the previous results for two samples in section \ref{sec:fixed-n} to be extended for all $n_a, n_b \in \mathbb{N}\times \mathbb{N}$.
This can be achieved by using elegant ``drop-in'' replacements for $\epsilon_n(\alpha)$ in equation \ref{eq:dkwepsilon}. \citet{darling804} propose using
\begin{equation}
\label{eq:dr_epsilon}
    \epsilon_n(\alpha) = \sqrt{\frac{(n+1)(2\log n - \log(\alpha(n^\star - 1)))}{n^2}}.
\end{equation}
This provides an $\alpha$-level confidence sequences that holds for all $n\geq n^{\star}$. \citet{szorenyi15} provide a result that holds $\forall n \in \mathbb{N}$ by using
\begin{equation}
    \epsilon_n(\alpha) = \sqrt{\frac{1}{2n}\log \frac{\pi^2 n^2}{3\alpha}}.
\end{equation}
Their results follows directly from a union bound of the DKWM inequality in \eqref{eq:dkwinequality} over $n$ and using $\sum_{n=1}^{\infty}1/n^2 = \pi^2/6$.
 \citet{Howard21} obtain a tighter confidence sequence with the same guarantee by using
\begin{equation}
    \begin{split}
        \label{eq:hr_epsilon}
        \epsilon_{n}(\alpha) &= 0.85  \sqrt{\frac{\log \log(e  n) + 0.8  \log(1612 / \alpha)}{n}}.\\
    \end{split}
\end{equation}
The authors use these results to derive sequential tests of $F_a(x) \leq F_b(x)$ and $F_a(x) = F_b(x)$ \cite{darling804, Howard21}. We choose to work with equation \eqref{eq:hr_epsilon} moving forward. We complement these results by contributing sequential $p$-values, stopping logic for accepting an approximately true null hypothesis, and an upper bound on the number of observations required for the test to stop.

\subsubsection{Sequential $p$-values}
A sequential $p$-value satisfies the following
\begin{equation}
    \mathbb{P}_{H_0}[p_{n_a,n_b} \leq \alpha \, \text{for some}(n_a, n_b) \in \mathbb{N}\times \mathbb{N}] \leq \alpha.
    \label{eq:seqpvalue}
\end{equation}
$p$-values are often requested by developers as a measure of strength against the null hypothesis and are also necessary inputs for procedures controlling false discovery rate \cite{Johari15}. The sequential $p$-value for testing $A \preccurlyeq B$ is equal to the root of \eqref{eq:dkwalbroot}, except that $\epsilon_{n_a, n_b}(\alpha)$ is replaced by its new definition in equation \eqref{eq:hr_epsilon}.
When $n_a = n_b = n$, the sequential $p$-value is given by
\begin{equation*}
    p_{n_a,n_b}^{\preccurlyeq}= \frac{3624}{e^{\frac{n \left(\frac{\|d_{n_a,n_b}^{+}\|_\infty/2}{0.85}\right)^2 - \log \log (en)}{ 0.8}}},
\end{equation*}
otherwise it is sandwiched between $p_{\max(n_a, n_b)}(\|d_{n_a,n_b}^{+}\|_\infty)$ and $p_{\min(n_a, n_b)}(\|d_{n_a,n_b}^{+}\|_\infty)$ where
\begin{equation*}
    p_{n}(d)= \frac{3624}{e^{\frac{n \left(\frac{d/2}{0.85}\right)^2 - \log \log (en)}{ 0.8}}},
\end{equation*}
and can be computed numerically via bracketed root finding algorithms. 
Sequential $p$-values for testing $A \succcurlyeq B$ and $A \overset{d}{=} B$ are obtained by replacing $\|d_{n_a, n_b}^{+}\|_\infty$ with $\|d_{n_a, n_b}^{-}\|_\infty$ and $\|d_{n_a, n_b}\|_\infty$, respectively. Practically, this permits the analysis to be performed as frequently as desired while maintaining coverage and type I error guarantees.
In particular, it permits the use of a data-dependent stopping rule. The null hypothesis can be rejected at the $\alpha$ level as soon as the corresponding sequential $p$-value falls below $\alpha$.

Observations are received in the experiment in no specific order from arms A and B, they may even arrive at different rates. Let $t=1,2,\dots$ be an enumeration of $\mathbb{N}\times\mathbb{N}$ in the order in which they occur. One can define a new sequential $p$-value by computing the running minimum $q_t = \min (q_{t-1}, p_{t})$ with $q_0=1$, which satisfies
\begin{equation}
    \mathbb{P}_{H_0}[q_t \leq \alpha \, \text{for some }\, t \in \mathbb{N}] \leq \alpha.
    \label{eq:seqqvalue}
\end{equation}
Confidence sequences on $\sup d_{a,b}(x)$, $\inf d_{a,b}(x)$ and $\|d_{a,b}\|_\infty$ can be constructed similarly, and are provided in Appendix \ref{app:sup_norm}.

\subsubsection{Accepting an Approximate Null Hypothesis}
One can continue to observe additional datapoints and tighten confidence bands on quantile and distribution functions forever. The more datapoints that are observed, the tighter these confidence bands become. For this reason, any difference in these functions is guaranteed to be revealed eventually i.e.\ the test is asymptotically power 1 \cite{Howard21}.

In practice, developers want to stop the test in a finite time, such as when they are satisfied that the distributions are ``similar enough''. Let $\tau$ denote a subjective tolerance specified by the developer, such that any departure from the null hypothesis by less than $\tau$ is practically irrelevant. We provide the developer a complementary stopping rule such that if the null is not already rejected, one can at least conclude with confidence that the difference is less than the desired tolerance.

For testing $F_a(x) \geq F_b(x)$ $\forall x \in \mathcal{X}$, we propose to stop when the time-uniform version of the upper confidence bound on $\sup d_{a,b}(x)$, $\sup d_{n_a, n_b}^{u}(\alpha, x)$, in corollary \ref{cor:infsupci} is less than $\tau$. 
This ensures with confidence at least $1-\alpha$ that the null is approximately true i.e.\ $F_a(x) \geq F_b(x) - \tau$ for all $x \in \mathcal{X}$. 
When testing $F_a(x) \leq F_b(x)$ $\forall x \in \mathcal{X}$, stopping when $\inf d_{n_a, n_b}^{l}(\alpha, x) > -\tau$ ensures with confidence at least $1-\alpha$ that $F_a(x) \leq F_b(x) + \tau$ for all $x \in \mathcal{X}$. 
Lastly, when testing $F_a = F_b$, stopping when $u_{n_a, n_b}(\alpha) < \tau$ ensures with confidence at least $1-\alpha$ that $|F_b(x) - F_a(x)| \leq \tau$ for all $x \in \mathcal{X}$.

\begin{table}
  \caption{Stopping Rules}
  \label{tab:stopping_rules}
  \begin{tabular}{ccl}
    \toprule
    Hypothesis & Reject & Approximate Accept\\
    \midrule
    $A \preccurlyeq B$ & $\sup d_{n_a, n_b}^{l}(\alpha, x) > 0$ & $\sup d_{n_a, n_b}^u(\alpha, x) < \tau$ \\
    $A \succcurlyeq B$ & $\inf d_{n_a, n_b}^{u}(\alpha, x) < 0$ & $\inf d_{n_a, n_b}^l(\alpha, x) > -\tau$ \\
    $A \overset{d}{=} B$ & $ l_{n_a, n_b}(\alpha)  >0$ & $u_{n_a, n_b}(\alpha) < \tau$\\
  \bottomrule
\end{tabular}
\end{table}

\subsubsection{Stopping Rules}
We provide a summary of the stopping rules for each null hypothesis in table \ref{tab:stopping_rules} (note that these definitions use the time-uniform version of $\epsilon_n(\alpha)$ in equation \eqref{eq:hr_epsilon}). The probability that the incorrect conclusion is drawn is at most $\alpha$. Similar "open-ended" versions of these can be found in \citet[Appendix B.2.]{Howard21}. The stopping logic for testing $A \preccurlyeq B$ is quite simple when stated in plain English. With a slight abuse of mathematical verbiage, $A \preccurlyeq B$ is rejected as soon as the confidence band on $d_{a,b}$ is strictly positive for some $x$, and approximately accepted when it is less than $\tau$ for all $x$. $A \overset{d}{=} B$ is rejected as soon as the confidence band on $d_{a,b}$ excludes zero for some $x$, and approximately accepted when it is within $-\tau$ and $\tau$ for all $x$. Note that $l_{n_a,n_b}>0$ is simply the logical \textit{OR} of the rejection criteria of tests $A \preccurlyeq B$ and $A \succcurlyeq B$, while $u_{n_a, n_b} < \tau$ is the logical \textit{AND} of the acceptance criteria.

Alternatively, one needn't reject immediately. A developer can continue observing datapoints until the confidence bands on the quantile or distribution functions are precise enough to satisfy their curiosity. In this case, the sequential $p$-value can be used as a final measure of evidence against the null hypothesis. One can also obtain a maximum number of observations required by the canary test by considering the number required to give a confidence band on $d_{a,b}$ of desired radius, which allows the developer to reason about meaningful differences, as discussed in section \ref{sec:samplesize}. When $n_a=n_b=n$, The number of observations required for a confidence band of radius $r$ satisfies $2\epsilon_{n}(\alpha/2) = r$, which can be solved numerically for $n$.

\subsection{Count Metrics}
\label{sec:renewal}
In general a datapoint $i$ from arm $k$ in a canary test is a 2-tuple $(x_{ki}, t_{ki})$ of a measurement $x_{ki}$ and an arrival timestamp $t_{ki}$. The measurement $x_{ki}$ is often continuous, like \textit{PlayDelay} in milliseconds, and $t_{ki}$ is a timestamp corresponding to when the measurement was taken or when it was received. In addition to testing hypotheses about the distribution of the measurement, it is also very relevant to test hypotheses about the timestamps. If the distributions of the measurements are equal among arms, but datapoints are received at a different rates, then this is also a cause for concern.

An extremely important example is when the metric corresponds to a particular error. In this example the measurement is simply an indicator that an error has occurred and the timestamp records when the error occurred. It would be highly alarming if the release candidate produced errors at a faster rate. Netflix carefully monitors playback errors which indicate a failure to play a title. If arm B experiences a higher volume of errors, then this is a strong signal of a regression. A fixed-$n$ approach is simply to test whether the number of errors produced in a window of time by arm B is greater or less than arm A. For this reason we refer to these as \textit{count metrics}.

The raw data for a count metric is simply an ordered sequence of timestamps. Consequently, we model this as a one-dimensional point process. \citet{kuo1996} model the point process of software failure times as a time-inhomogeneous Poisson point process, or equivalently, the epoch's of failure times as a time-inhomogeneous exponential process. We would prefer, however, not to make such strong assumptions. Instead, we model the sequence of timestamps as a general renewal process.

A renewal process is a stochastic process in time with waiting times between successive timestamps (epochs) drawn i.i.d.\ from a \textit{renewal distribution}. In contrast to the Poisson point process assumption, no further assumptions are made about the renewal distribution. Suppose the canary test begins at time $t=0$ and the current time is $\mathcal{T}$. During this time, a sequence of datapoints has arrived in arm A at times $t_{a1}, t_{a2}, \dots t_{an_a}$. Let $G_a$ be the renewal distribution of arm A. The likelihood for $G_a$ given the observed times is then
\begin{equation}
\begin{split}
        p((t_{a1}, t_{a2}, \dots, t_{an_a})|G_a) = &\prod_{i=1}^{n-1} g_a(t_{ai+1} - t_{ai})\\
        &g_a(t_{a1})(1-G_a(\mathcal{T}-t_{an_a}),
\end{split}
\end{equation}
where $g_a$ is the Radon–Nikodym derivative of $G_a$ with respect to Lebesgue measure. Associated with each event $t_{ai}$ can be an additional measurement $x_{ai}$ modelled as a random variable from the distribution $F_a$. Our proposal is to sequentially test for differences in both the measurement distributions $F_a$ and $F_b$ as well as the renewal distributions $G_a$ and $G_b$. Testing for differences in the renewal distribution is as simple as feeding the time differences $(t_{a2} - t_{a1}), (t_{a3} - t_{a2}), \dots $ for arms A and similarly for B into the procedure described in section \ref{sec:sequential}.

\section{Case Studies}
\label{sec:casestudies}
In this section, we describe two case studies which demonstrate the approach described above in the context of canary testing client applications at Netflix. These tests are based on a controlled experiment where a randomized group of users are assigned to receive the release candidate while a control group receive the existing version of the client application. These tests are configured to run until the null is rejected or approximately accepted. Towards this end, our approach was deployed as a quality gate in the existing software delivery pipeline for the client applications under evaluation. These quality gates were configured to alert developers when a regression was detected with the release candidate.

\subsection{Increase in \textit{PlayDelay}}
\begin{figure}[h]
\includegraphics[width=8cm]{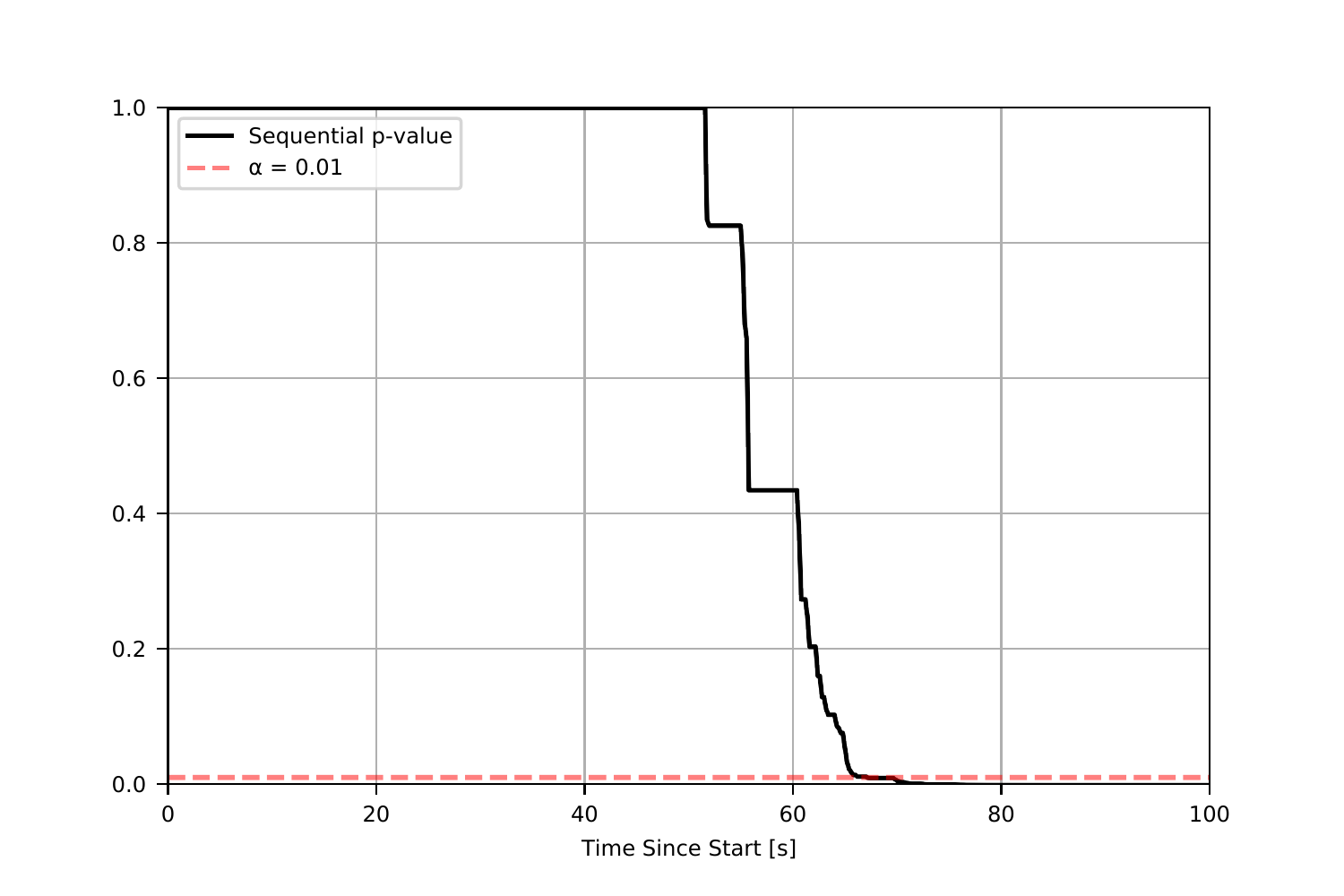}
\caption{Sequential $p$-value $q_t$ from equation \eqref{eq:seqqvalue} for \textit{PlayDelay}.}
\label{fig:pd_pvalue}
\end{figure}

In the following case study, we show how the sequential methodology successfully detected a performance regression in \textit{PlayDelay} and prevented the release candidate from reaching the production environment. The data is taken from a canary test in which the behavior of the client application was modified. The null hypothesis was that observations of \textit{PlayDelay} in the release candidate should be stochastically less than or equal to observations in the existing version ($H_0: F_b \geq F_a$). Figure \ref{fig:pd_pvalue} shows the sequential $p$-value for this hypothesis as a function of time since the beginning of the canary test. The $p$-value falls below $0.01$ after approximately 65 seconds.

\begin{figure}[h]
\includegraphics[width=8cm]{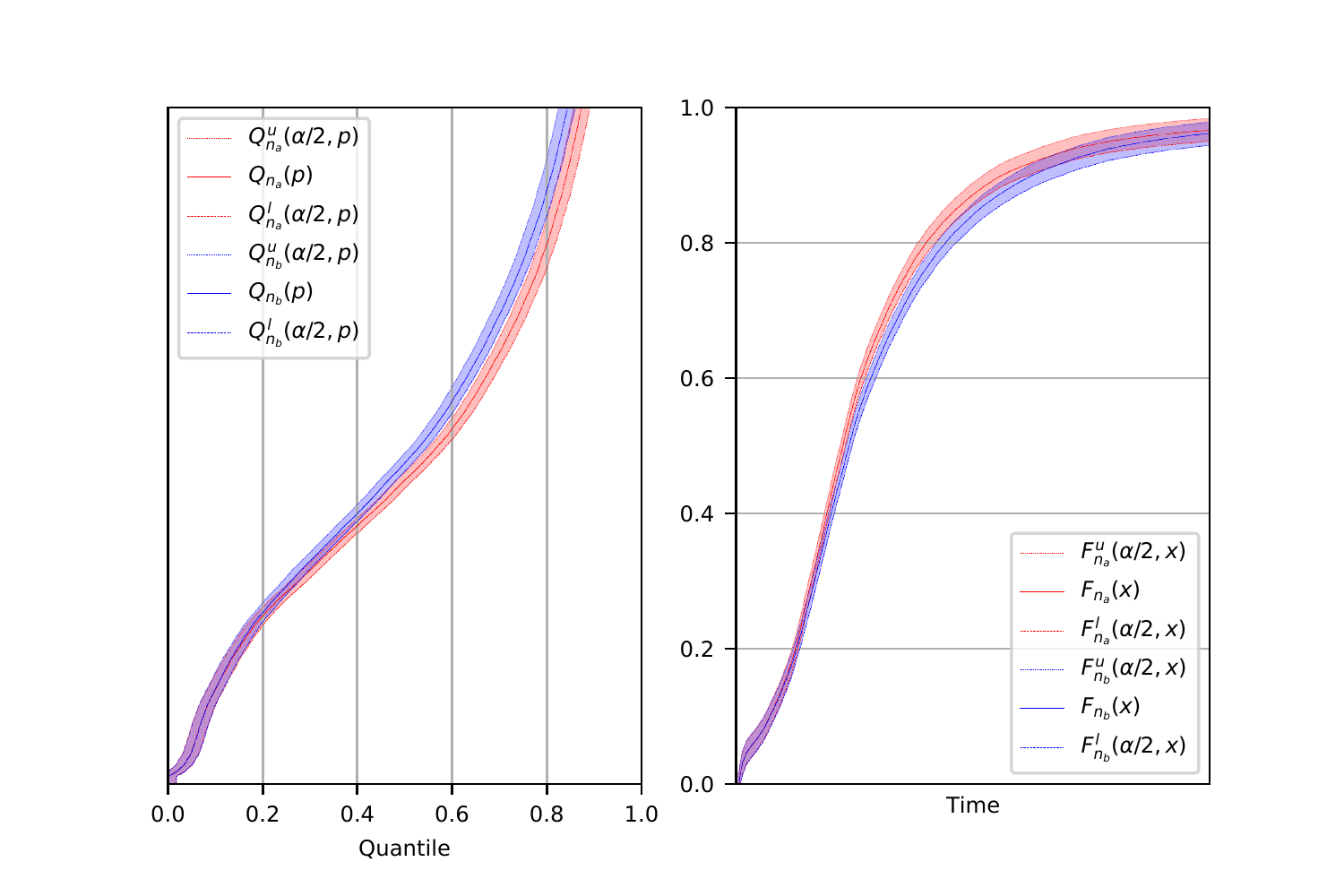}
\caption{Empirical quantile (left) and distribution (right) functions of \textit{PlayDelay} with 0.995 confidence bands from equations \eqref{eq:cdfband} and \eqref{eq:quantileband} at time t=100 seconds. Ticks omitted for confidentiality.}
\label{fig:pd_quantile}
\end{figure}

Figure \ref{fig:pd_quantile} shows that the $0.995$ confidence bands for the quantile functions of arms A and B fail to intersect for many quantiles. While no quantiles are significantly lower in arm B, many are significantly greater, revealing a small but significant increase in \textit{PlayDelay}. This is perhaps easier to see by considering the $0.99$ confidence band on the difference function in figure \ref{fig:pd_quantile_difference}.
The median value of \textit{PlayDelay} increased by at least 11 and at most 255 milliseconds in arm B, while the 75-percentile increased by at least 51 and at most 635 milliseconds (based on a 0.99 confidence interval).  If the developer wishes to get more precise estimates and tighter confidence intervals, they can continue the canary test without sacrificing coverage guarantees due to the time-uniform nature of these confidence sequences.

\subsection{Drop in \textit{Successful Play Starts}}
While \textit{PlayDelay} captures the time taken by successful playbacks, it fails to capture unsuccessful playbacks (playback failures). To complement \textit{PlayDelay}, Netflix also closely monitors the number of successful play starts. Simply, a \textit{Successful Play Start (SPS)} event is sent to the central logging system by the client application whenever a title begins streaming after the viewer hits play, logging the successful start of the title. A significant drop in \textit{SPS} events between arms therefore implies a significant increase in playback failures, and is an important metric for detecting software regressions. The following case study uses real data from a canary test where a software bug caused playback to fail for multiple devices.

The number of \textit{SPS} events is an example of a \textit{count metric} described in section \ref{sec:renewal}. Consequently, we look at the time-differences between observed events (epochs) to study differences in the renewal distributions. In this example we test $F_a = F_b$, where $F_i$ denotes the renewal distribution for arm $i$.

\begin{figure}[h]
\includegraphics[width=8cm]{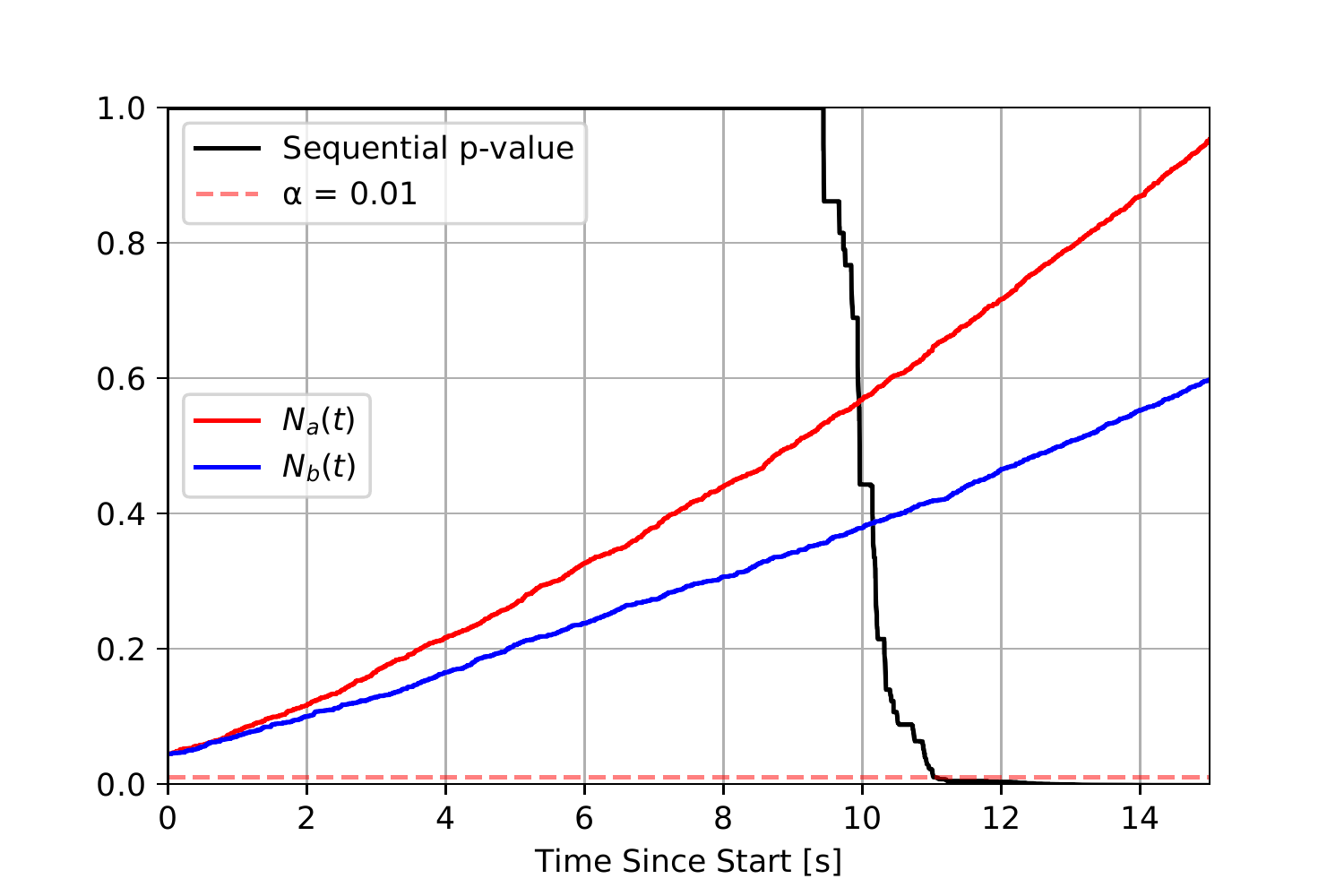}
\caption{(Left Axis) Sequential $p$-value $q_t$ from equation \eqref{eq:seqqvalue} for \textit{SPS}. (Right Axis) Cumulative counts of \textit{SPS} events over time for arms A and B. Ticks omitted for confidentiality.}
\label{fig:sps_counts}
\end{figure}

Let the counting process $N_i(t)$ be defined as the number of events observed in the interval $(0,t]$ for arm $i$. These are shown on the right axis of figure \ref{fig:sps_counts}. Clearly there are fewer \textit{SPS} events per unit time for arm B running the newer client application, suggesting a bug in the release candidate which prevents some users from streaming.

The left axis of figure \ref{fig:sps_counts} shows the sequential $p$-value, which falls below $\alpha=0.01$ in a mere $11.08$ seconds. 
From a different perspective, figure \ref{fig:sps_supnorm} shows the confidence sequence on $\|d_{a,b}\|_\infty$. Note the sequential $p$-value falls below $\alpha=0.01$ at the same time the confidence sequence excludes zero, when $l_{t}(0.01) >0$.

\begin{figure}[h]
\includegraphics[width=8cm]{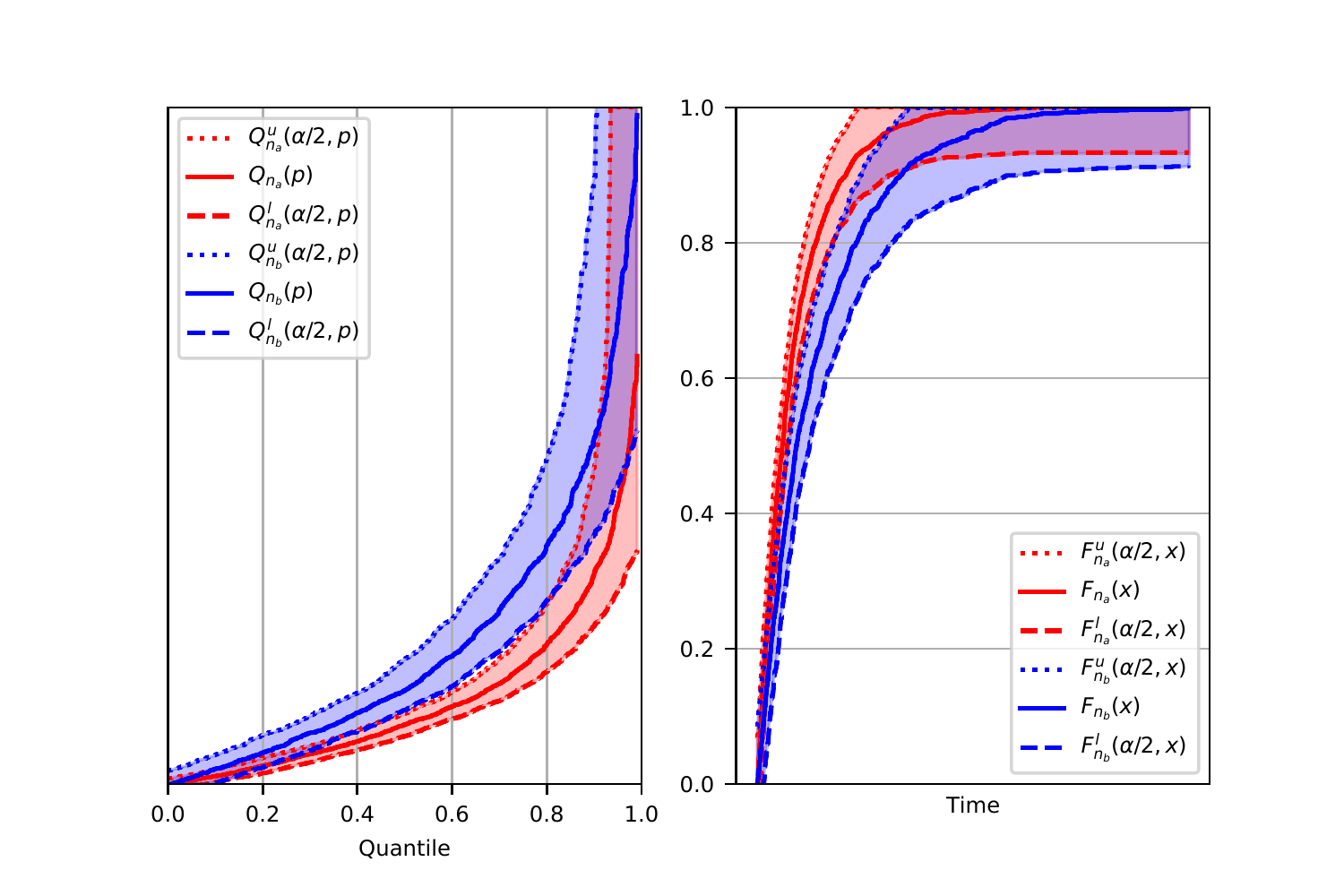}
\caption{Empirical quantile (left) and distribution (right) functions of epochs of \textit{SPS} with $0.995$ confidence bands from equations \eqref{eq:cdfband} and \eqref{eq:quantileband} at t=15 seconds. Ticks omitted for confidentiality.}
\label{fig:sps_quantile}
\end{figure}

Figure \ref{fig:sps_quantile} shows the empirical quantile and distribution functions with $0.995$ confidence bands at 15 seconds into the canary test. It is clear that many quantiles are shifted toward larger values, implying that the distribution of time-differences is shifted toward larger values and that \textit{SPS} events are arriving less frequently.

We regard the confidence sequences on the quantile and distribution functions, as well as their differences, to be complementary to the automated stopping logic. They allow the developer to be kept in the loop as the canary test progresses and provide them insight into what would otherwise be a black-box system. Due to the time-uniform guarantees these can be visualized at any time without concern of peeking - the coverage guarantee holds for all time.

\section{Discussion}
\label{sec:discussion}
The case studies presented above demonstrate the utility of our approach in rapidly detecting regressions. In both cases, our approach was successful in alerting developers to an issue within seconds. This is in contrast to the existing approach, based on fixed-$n$ tests, which would have taken upwards of 30 minutes or longer to detect the regression. This would have exposed the treatment population to a degraded experience for a prolonged period of time. 

In practice, it can be tempting to apply fixed-$n$ tests for continuous monitoring. In our own experience, we have observed developers repeatedly applying these tests to quickly detect regressions. However, as demonstrated in Appendix \ref{sec:simulations}, the result of doing this is beyond just theoretical, i.e., it can result in an unacceptable number of false positives. This can erode trust in the canary test and impact deployment velocity as developers search for sources of regressions that do not exist.

For count metrics our methodology makes an assumption that the distribution of time-differences between successive observations is stationary. We have observed this to be a reasonable assumption for metrics with a high rate of events, as evinced in figure \ref{fig:sps_counts}, because the run-time of the test is short. Count metrics for which data arrives very slowly, such as \textit{SPS} for a rarely used device, take longer to reject or approximately accept the null. Therefore, over the course of a long-running canary test, it is likely that the assumption of a stationary renewal distribution is violated due to time-varying usage patterns of the Netflix application. Despite this, it is not clear to what extent the type I error guarantees of this system would degrade, given that it affects both arms equally. An alternative approach could be a sequential multinomial test to compare the proportion of counts across arms, which remains valid even in the presence of time-inhomogeneity \cite{srm}.

\section{Conclusion}
\label{sec:conclusions}
This paper presents an approach to canary testing that has successfully detected performance regressions and bugs in software deployments at Netflix. The novel contributions of this approach are the formulation of regressions in terms of stochastic orderings and the sequential testing framework. Testing hypotheses of stochastic order enables developers to assign an undesirable direction to changes in a metric distribution beyond simply comparing the means. When no direction is available or appropriate, any difference in the distribution can be tested.

The sequential methodology permits canary tests to be \textit{continuously monitored}, while retaining strict type I error guarantees. This allows automated stopping logic to be implemented for the canary test, removing the burden on the developer to monitor the release manually and freeing up time for them to return other tasks. The end effect is that bugs and performance regressions are detected rapidly, prevented from reaching the production environment and subsequent experimental units are saved from a degraded experience. Near-immediate feedback is given to the developer, allowing them to iterate quickly and remedy the issue.

\begin{acks}
The authors would like to thank Minal Mishra, Toby Mao, Yanjun Liu, and Martin Tingley for their contributions. The authors would also like to thank the reviewers for their comments and suggestions.
\end{acks}

\bibliographystyle{ACM-Reference-Format}
\bibliography{references}

\clearpage
\appendix
\section*{Appendix}

\section{Confidence Band on $|d_{a,b}|$}
\label{app:absdiff}
\begin{figure}[h]
\includegraphics[width=8.5cm]{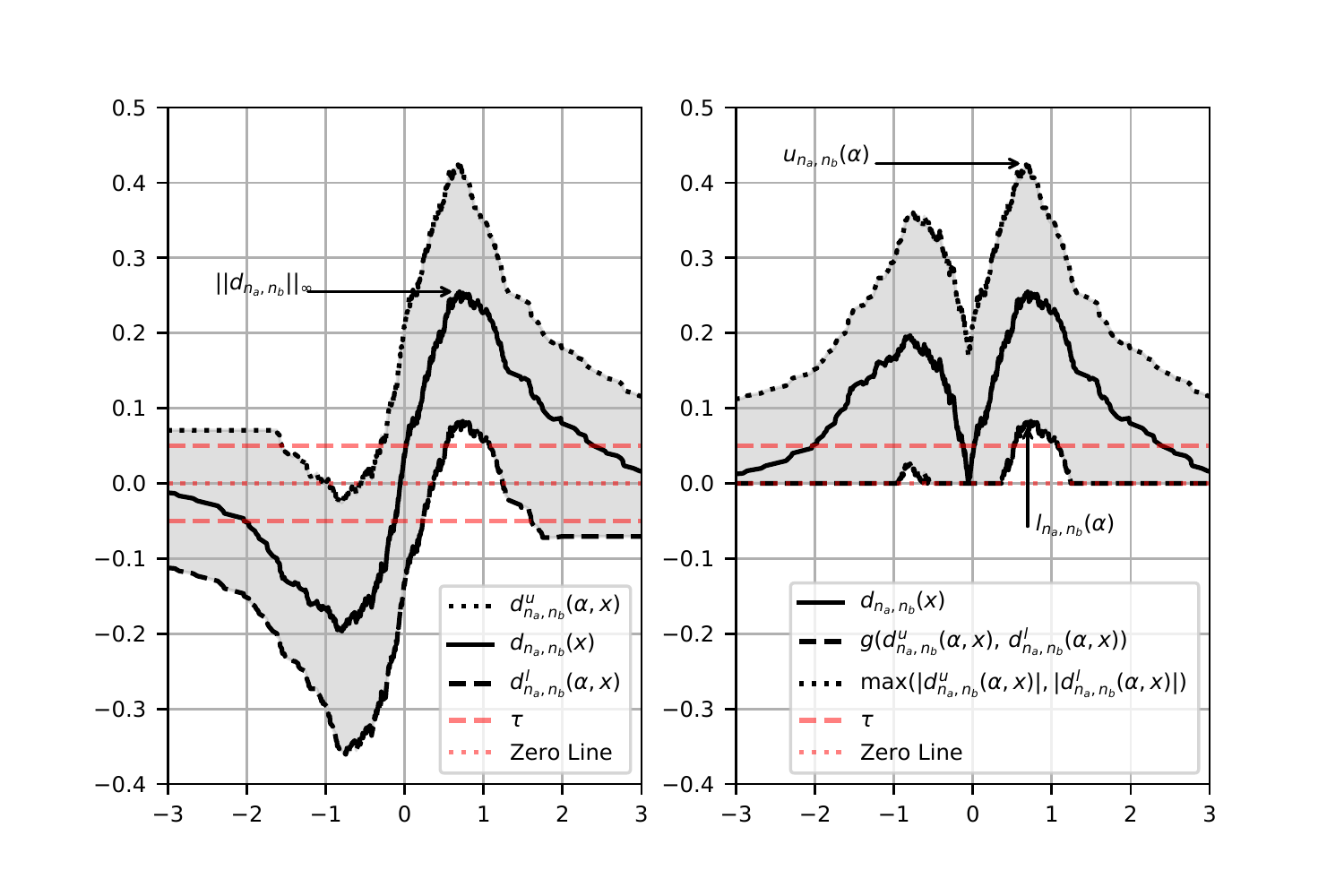}
\caption{Empirical difference (left) and absolute difference (right) functions with $1-\alpha$ confidence bands from equations \eqref{eq:diffconfband} and \eqref{eq:absdiffconfband} using the same data as in figure \ref{fig:dkwdistquants}.}
\label{fig:dkwdiff}
\end{figure}

The confidence band on $d_{a,b}$ can be used to derive a confidence band on $|d_{a,b}|$, this is illustrated in figure \ref{fig:dkwdiff}. Note that the image of $[l,u]$ under $x \rightarrow |x|$ is $[0, \max(|l|,|u|)]$ if $0 \in [l,u]$ or $[\min (|l|, |u|), \max(|l|,|u|)]$ otherwise. Define
\[
    g(x,y)= 
\begin{cases}
    \min(|x|,|y|),& \text{if } \text{sign}(x) = \text{sign}(y)\\
    0,              & \text{otherwise},
\end{cases}
\]
then
\begin{equation}
\begin{split}
     \mathbb{P}[
     & g(d_{n_a,n_b}^u(\alpha,x),\,d_{n_a,n_b}^l(\alpha,x)) \leq\\
     & |d_{a,b}(x)| \leq \\
    & \max(|d_{n_a,n_b}^u(\alpha,x)|,\, |d_{n_a,n_b}^l(\alpha,x)|)\\
     &\forall x \in \mathcal{X}] \geq 1-\alpha,
\end{split}
\label{eq:absdiffconfband}
\end{equation}

\section{Proof of Theorem \ref{thm:a<=b}}
\label{app:a<=b}
The null hypothesis $F_a(x) \geq F_b(x) \, \forall x \in \mathcal{X}$ is equivalent to $d_{a,b}(x) \leq 0 \, \forall x \in \mathcal{X}$. It can be rejected at the $\alpha$-level if there exists an $x \in \mathcal{X}$ such that $d_{n_a,n_b}^{l}(\alpha,x) > 0$ i.e.\ when the lower bound on the confidence interval of $F_b(x)$ is greater than the upper bound on the confidence interval of $F_a(x)$. This is true if and only if $\sup \lbrace d_{n_a,n_b}^{l}(\alpha, x): x \in \mathcal{X} \rbrace > 0$. 

To obtain a $p$-value we seek the smallest such $\alpha$-level test that rejects the null i.e.
\begin{equation}
    p_{n_a,n_b}^{\preccurlyeq} = \inf \left\{ \alpha \in [0,1]: \underset{x\in\mathcal{X}}{\sup}d_{n_a,n_b}^{l}(\alpha, x) > 0 \right\}
\end{equation}
Let $d_{n_a,n_b}^{+}(x) = \max (d_{n_a,n_b}(x), 0)$ denote the positive part of the empirical difference function.
If there exists an $x \in \mathcal{X}$ such that $d_{n_a,n_b}(x) > 0$, this set is not empty, and $p_{n_a,n_b}^{\preccurlyeq}$ is the root of the following equation
\begin{equation}
    f(\alpha) =  \|d_{n_a,n_b}^{+}\|_\infty - \epsilon_{n_a,n_b}(\alpha) 
    \label{eq:root}
\end{equation}
where $\epsilon_{n_a,n_b}(\alpha) = \epsilon_{n_a}(\alpha/2)+ \epsilon_{n_b}(\alpha/2)$. The intuition here comes from finding the largest confidence band radius such that the test still rejects. This critical value occurs when the confidence band radius is equal to the largest positive value of $d_{n_a, n_b}$ over the domain $\mathcal{X}$.
For any $\alpha < p_{n_a,n_b}$, $\sup d_{n_a,n_b}^{l}(\alpha, x) \leq 0$, which implies that $d_{n_a,n_b}^{l}(\alpha,x) \leq 0$ $\forall x \in \mathcal{X}$ and the null hypothesis is not be rejected.

\section{Additional Figures}
\begin{figure}[h]
\includegraphics[width=8cm]{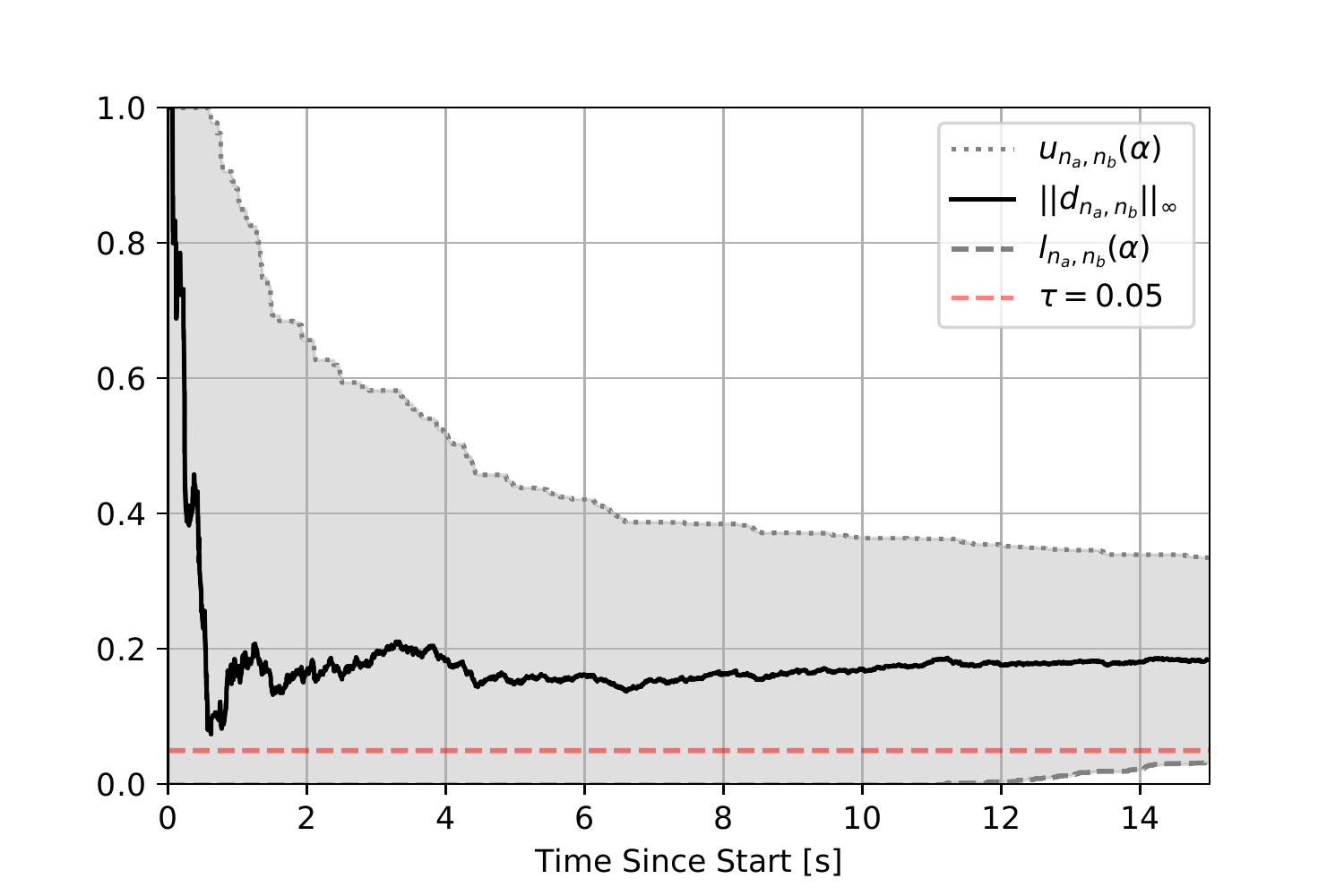}
\caption{Confidence sequence on the sup-norm, as in equation \eqref{eq:sup_norm_confidence_sequence}, of the difference between renewal distribution functions for \textit{SPS} case study.}
\label{fig:sps_supnorm}
\end{figure}

\begin{figure}[h]
\includegraphics[width=8cm]{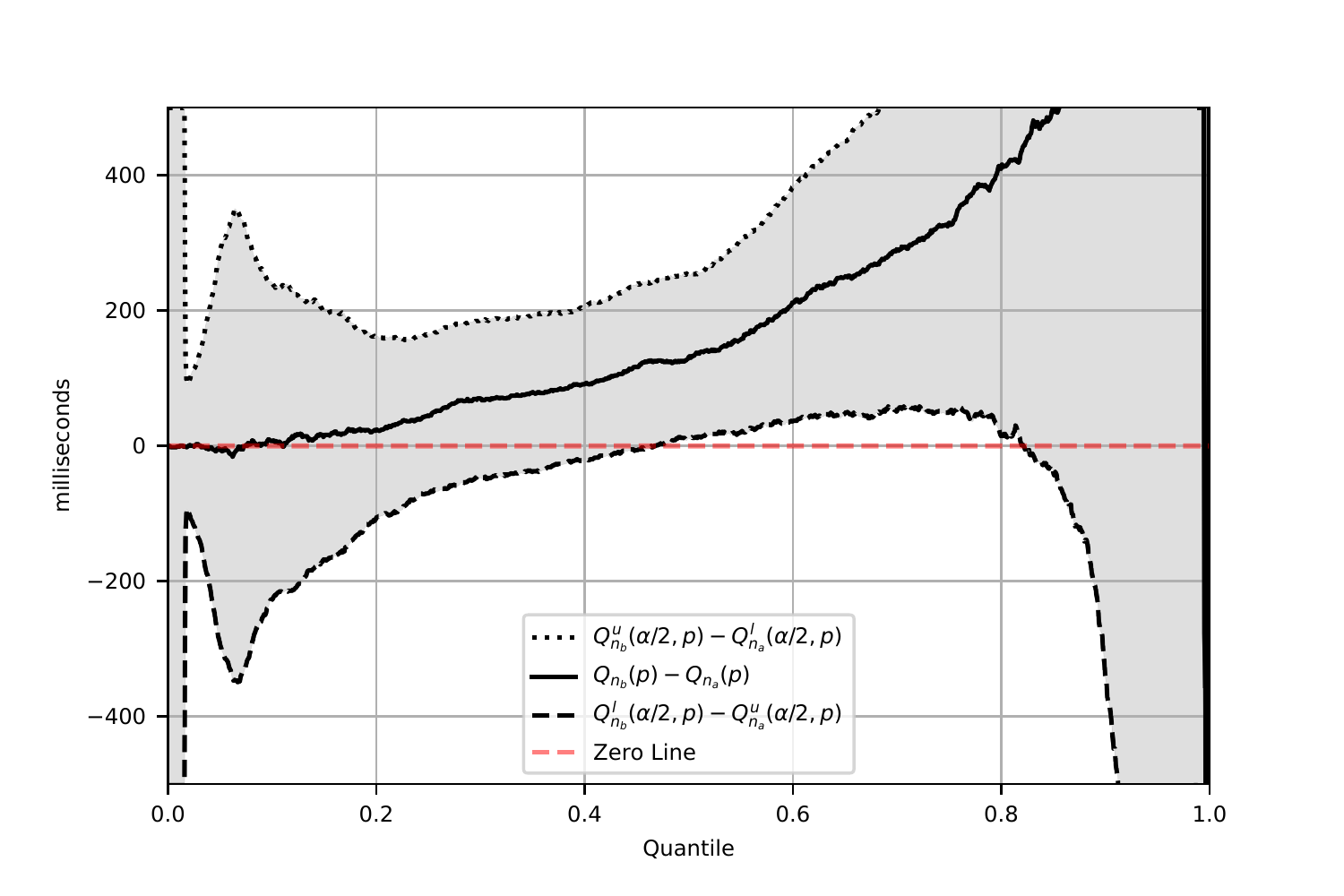}
\caption{$0.99$ confidence band on $Q_b-Q_a$ for \textit{PlayDelay} at time t=100 seconds.}
\label{fig:pd_quantile_difference}
\end{figure}

\section{Confidence Sequences on $\|d_{a,b}\|_\infty$}
\label{app:sup_norm}
One can define confidence sequences for $\sup d_{a,b}(x)$, $\inf d_{a,b}(x)$ and $\|d_{a,b}\|_\infty$ by taking running intersections, e.g.
\begin{equation}
    \mathbb{P}\left[\underset{x\in\mathcal{X}}{\sup} d_{a,b}(x) \in  \bigcap_{t=1}^{\infty} \left[
    \underset{x\in\mathcal{X}}{\sup}d_{t}^{l}(\alpha, x), \underset{x\in\mathcal{X}}{\sup}d_{t}^{u}(\alpha,x)\right]\right] \geq 1-\alpha
\end{equation}
\begin{equation}
    \mathbb{P}\left[\underset{x\in\mathcal{X}}{\inf} d_{a,b}(x) \in \bigcap_{t=1}^{\infty} \left[
    \underset{x\in\mathcal{X}}{\inf}d_{t}^{l}(\alpha, x), \underset{x\in\mathcal{X}}{\inf}d_{t}^{u}(\alpha,x)\right]\right] \geq 1-\alpha
\end{equation}
\begin{equation}
    \mathbb{P}\left[ \|d_{a,b}\|_\infty \in \bigcap_{t=1}^{\infty} [l_t(\alpha), u_t(\alpha)]\,\right] \geq 1-\alpha.
    \label{eq:sup_norm_confidence_sequence}
\end{equation}

\section{Simulation Study}
\label{sec:simulations}
In this section, we present a simulation which demonstrates the advantages of the sequential test compared to fixed-$n$ methods. To illustrate the advantage of sequential over fixed-$n$ methods, we examine the empirical type I error probabilities under continuous monitoring, that is, performing a significance test after every datapoint. We generate 100 simulations where each simulation generates independent streams of i.i.d.\ $\text{Gamma}(10,10)$ random variables for each arm. After every new pair of observations a significance test configured at the $\alpha = 0.05$ level is performed. If the $p$-value is less than 0.05, the null hypothesis is rejected and the stopping time is recorded, otherwise a new pair of observations is sampled from arms A and B. The tests compared the fixed-$n$ Kolmogorov-Smirnoff and Mann-Whitney tests to the sequential test in section \ref{sec:sequential}. The random number generator is seeded consistently, so that each stream of random variables is identical for each test. To ease the computation, the simulations are terminated when 5000 observations are sampled from each arm. 

\begin{figure}
\includegraphics[width=8.5cm]{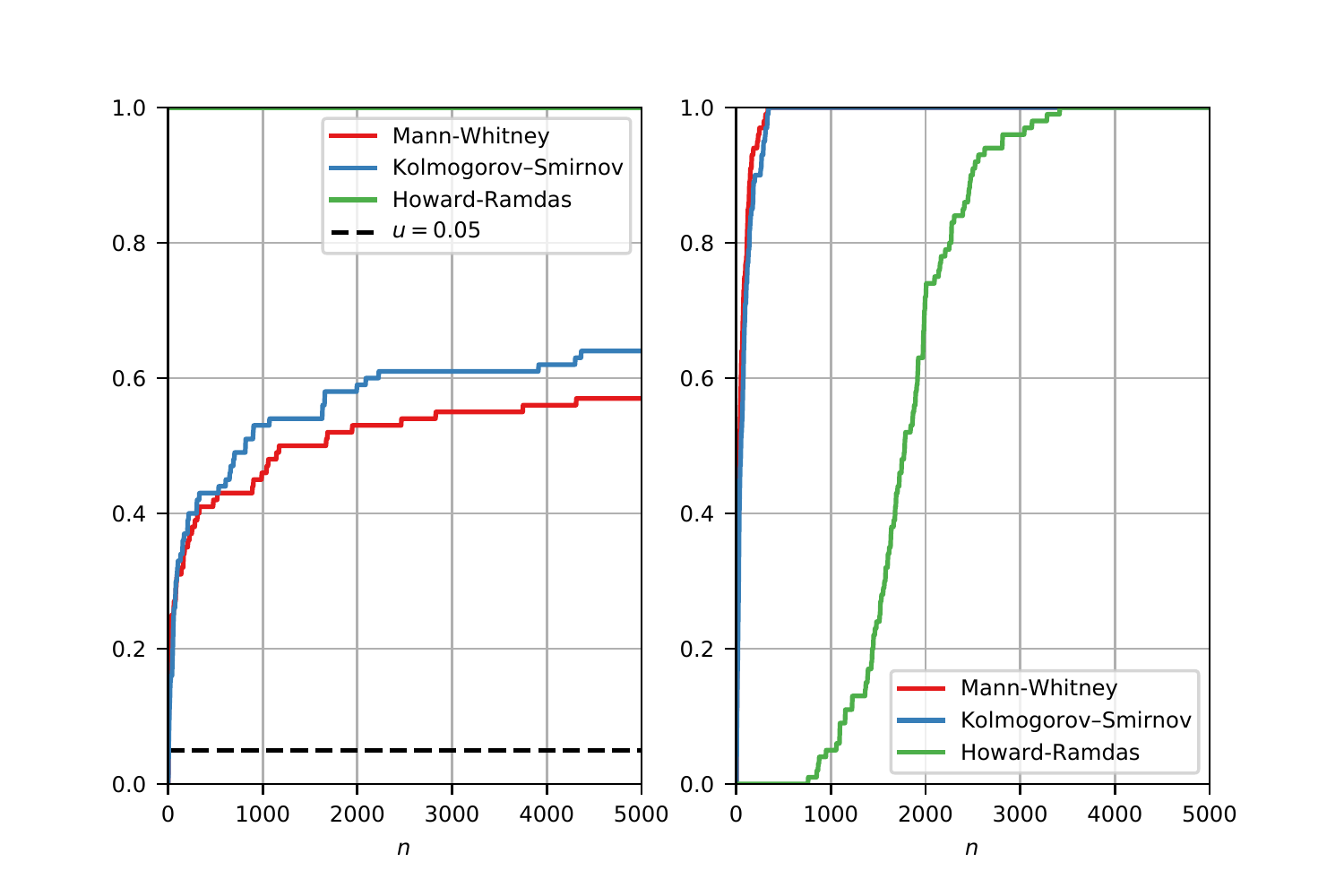}
\caption{Empirical distribution functions of the stopping times based on 100 observations for tests configured at the $\alpha = 0.05$ level testing the null hypothesis $F_a = F_b$. (Left) Under the null hypothesis: distributions for arms A and B are $\text{Gamma}(10,10)$. (Right) Under the alternative hypothesis: distributions for arms A and B are $\text{Gamma}(10,10)$ and $\text{Gamma}(10,11)$ respectively.}
\label{fig:stopping_time}
\end{figure}

The left of figure \ref{fig:stopping_time} clearly visualizes the problem with continuous monitoring for the Kolmogorov-Smirnoff and Mann-Whitney tests, which resulted in 64 and 57 false positives respectively. It also shows that no type I errors resulted from the sequential procedure. This can be explained by the fact that the sequential procedure must control the type I error probability \textit{for all} $(n_a,n_b) \in \mathbb{N}\times\mathbb{N}$, and so one expects the type I error probability in a smaller, finite window of time to be less than if it were to run indefinitely. The right of figure \ref{fig:stopping_time} shows the empirical probability of correctly rejecting the null by $n$ in simulations where the null is incorrect.

\end{document}